\documentstyle[preprint,aps,eqsecnum,floats,amssymb]{revtex}
\def \be{\begin{equation}}
\def \ee{\end{equation}}
\def \bea{\begin{eqnarray}}
\def \eea{\end{eqnarray}}
\def \ben{\begin{enumerate}}
\def \een{\end{enumerate}}
\def \bk{\bar{B}\to K^-\p^+ e^-e^+}
\def \bkc{B\to K^+\p^- e^-e^+}
\def \bp{\bar{B}\to \p^-\p^+ e^-e^+}
\def \bpc{B\to \p^-\p^+ e^-e^+}
\def \sd{\text{SD}}
\def \weak{\text{w}}

\def \heff{H_{\text{eff}}}
\def \Im{{\text{Im}}\,}
\def \Re{{\text{Re}}\,}

\def \GeV{{\text{GeV}}}

\def \av#1{\left\langle #1\right\rangle}
\def \bm{\boldmath}
\def \braket#1#2#3{\langle #1|#2| #3\rangle}
\def \cp{\mathrm{CP}}
\def \dis{\displaystyle}
\def \ea{{\it et al.}}
\def \eq#1{Eq.~(\ref{#1})}
\def \eqs#1#2{Eqs.~(\ref{#1})--(\ref{#2})}

\def \nnu{\nonumber}
\def \ol#1{\overline{#1}}

\def \rf{Ref.~\cite}

\def \sec#1{Sec.~\ref{#1}}
\def \Vec#1{{\bf{#1}}}
\def \cseff {c_7^{\text{eff}}}
\def \ceff {c_9^{\text{eff}}}

\def \a{\alpha}
\def \b{\beta}
\def \D{\Delta}
\def \g{\gamma}
\def \G{\Gamma}
\def \d{\delta}
\def \epsi{\epsilon}

\def \l{\lambda}

\def \m{\mu}
\def \n{\nu}

\def \p{\pi}
\def \r{\rho}
\def \s{\sigma}

\def\ib#1#2#3{{\it ibid.\/}~{\bf#1} (19#2) #3}

\def\np#1#2#3{{\it Nucl.~Phys.\/}~{\bf B#1} (19#2) #3}

\def\pl#1#2#3{{\it Phys.~Lett.\/}~{\bf B#1} (19#2) #3}

\def\pr#1#2#3{{\it Phys.~Rev.\/}~{\bf #1} (19#2) #3}
\def\prd#1#2#3{{\it Phys.~Rev.\/}~{\bf D#1} (19#2) #3}
\def\prl#1#2#3{{\it Phys.~Rev.~Lett.\/}~{\bf #1} (19#2) #3}

\def\rmp#1#2#3{{\it Rev.~Mod.~Phys.\/} {\bf #1} (19#2) #3}

\begin{document}
\preprint{\setlength{\baselineskip}{1.5em}
\vbox{\vspace{-1cm}
\hbox{FISIST/5-99/CFIF}
\hbox{IMSc-99/05/17}
\hbox{PITHA 99/15}
\hbox{hep-ph/9907386}
\hbox{July 1999}}}
\draft
\title{Angular Distribution and \bm$\cp$ Asymmetries in
the Decays $\bbox{\bk}$ and 
$\bbox{\bp}$}
\author{\sc Frank  Kr\"uger\thanks{E-mail address: krueger@gtae3.ist.utl.pt}}
\address{Centro de F\'{\i}sica das Interac\c{c}\~{o}es Fundamentais (CFIF),
Departamento de F\'{\i}sica,  Instituto Superior T\'ecnico,  
Av. Rovisco Pais,  1049-001 Lisboa, Portugal}
\author{\sc Lalit  M.~Sehgal\thanks{E-mail address: sehgal@physik.rwth-aachen.de}}
\address{Institut f\"ur Theoretische Physik (E), RWTH Aachen, D-52056 Aachen, Germany} 
\author{\sc Nita  Sinha\thanks{E-mail address: nita@imsc.ernet.in} 
and Rahul  Sinha\thanks{E-mail address: sinha@imsc.ernet.in}}
\address{Institute of Mathematical Sciences, Taramani, Chennai  600113, India}
%
%
\maketitle
\begin{abstract}
The short-distance Hamiltonian describing $b\to s(d) e^-e^+$ in the standard model is used  to obtain the decay spectrum of $\bk$ and $\bp$, assuming the $K\p$ and $\p\p$ systems to be the decay products of $K^*$ and $\r$ respectively. 
Specific features calculated are (i) angular distribution of $K^-$ (or $\p^-$) in the 
$K^-\p^+$ (or $\p^-\p^+$) centre-of-mass (c.m.) frame; (ii) angular distribution of $e^-$ in the $e^-e^+$ c.m.~frame; and (iii) the correlation between the meson and lepton planes. We also derive $\cp$-violating 
observables obtained by combining the above decays with the conjugate 
processes $\bkc$ and $\bpc$. 
\end{abstract}
\pacs{PACS number(s): 11.30.Er, 13.20.He}
%
%
\section{Introduction}\label{intro}
The purpose of this paper is to investigate the angular distribution of the decays $\bk$ and $\bp$, when the $K\p$ and $\p\p$ systems are the decay products of $K^*$ and $\r$ respectively. The aim is to derive the detailed consequences of the effective short-distance Hamiltonian describing the four-fermion interaction $b\to s(d) e^-e^+$. This work may be regarded as an extension of previous studies of the exclusive processes $\bar{B}\to K^* (\r) e^-e^+$ 
\cite{lit:btokstar,fklms:exc}
that were limited to the kinematical variables $s_l$ (the invariant mass of the lepton pair) and $\cos\theta_l$ (the angular distribution of $l^-$ in the 
$l^-l^+$ c.m.~system). The additional information we provide is the distribution in $\cos\theta_P$, 
where $\theta_P$ is the angle of the $K^-$ ($\p^-$) in the $K^-\p^+$ ($\p^-\p^+$) 
c.m.~frame, and the dependence on the angle $\phi$ between the $e^-e^+$ and 
$K^-\p^+$ or $\p^-\p^+$ planes. This information is sensitive to the polarization state of the vector meson $K^*$ ($\r$), and thus provides a new probe of the effective Hamiltonian.

An important aspect of the calculation is the prediction of $\cp$-violating observables that can be obtained by combining information from $B$ and $\bar{B}$ decays. These 
observables probe a term in the Hamiltonian proportional to $\Im (V_{ub}^{}V_{us}^*/V_{tb}^{}V_{ts}^*)$ (in the case of $B\to K^*$) and $\Im (V_{ub}^{}V_{ud}^*/V_{tb}^{}V_{td}^*)$ (in the case of $B\to \r$). While the numerical estimates of these asymmetries in the standard model turn out to be small, the formalism we present is also applicable to more general 
Hamiltonians transcending the standard model. The dependence on the variable $\phi$ can 
be an especially useful probe of $\cp$ violation, as has been demonstrated in the analogous case of $K_L\to \p^+\p^- e^+ e^-$ \cite{kdecay:lalitetal,kdecay:exp}. 

\section{Matrix element}\label{matrixelement}
We are concerned with the matrix element of the decay $\bar{B}(p)\to P  (k_1) 
P'  (k_2)l^+ (q_1) l^- (q_2)$, where $PP'= K^-\p^+$ or $\p^-\p^+$ and 
$l=e, \m$. 
Introducing the linear combinations \cite{pais,dafne}
\be\label{fourmomenta}
k = k_1+k_2,\quad  K=k_1-k_2,  \quad q = q_1+q_2,  \quad Q = q_1-q_2,
\ee
the short-distance Hamiltonian for $b\to f l^+l^-$ ($f=s$ or $d$) is
\cite{fklms:exc,bmm,review}  
\bea
\heff&=&\frac{G_F\a}{\sqrt{2}\p}V_{tb}^{}V_{tf}^*\bigg\{\ceff(\bar{f}\g_{\m}P_L b )\bar{l}\g^{\m}l+ c_{10}( \bar{f}\g_{\m}P_L b )\bar{l}\g^{\m}\g^5 l\nnu\\
& &\mbox{}-2\cseff \bar{f} i \sigma_{\m\n}\frac{q^{\n}}{q^2}(m_b P_R + m_f P_L)b\, \bar{l}\g^{\m} l\bigg\},
\eea
where  $P_{L,R}= (1\mp \g_5)/2$ and  
\be
\cseff= -0.315, \quad c_{10}= -4.642,
\ee
\bea\label{wilsonc9}
\ceff&=&c_9 +  (3 c_1 + c_ 2)\Big\{g(m_c,s_l) + \l_u\Big[g(m_c,s_l)
-g(m_u,s_l)\Big]\Big\}+ \cdots\nnu\\
&=&4.224 + 0.361\Big[(1+\l_u)g(m_c,s_l)-\l_u g(m_u,s_l)\Big]+ \cdots,
\eea
with 
\be
\l_u \equiv \frac{V_{ub}^{}V_{uf}^*}{V_{tb}^{}V_{tf}^*}, 
\ee
\bea\label{loopfunc}
\lefteqn{g(m_i,s_l)=-\frac{8}{9}\ln(m_i/m_b)+\frac{8}{27}+\frac{4}{9}y_i
-\frac{2}{9}(2+y_i)\sqrt{|1-y_i|}}\nnu\\[.7ex]
&&\times\left\{
\Theta(1-y_i)\left(\ln\left(\frac{1 + \sqrt{1-y_i}}{1 - \sqrt{1-y_i}}\right)-i\p\right)
+ \Theta(y_i-1) 2\arctan\frac{1}{\sqrt{y_i-1}}\right\},
\eea
where $s_l\equiv q^2$ and $y_i = 4 m_i^2/s_l$. The ellipses in the above
represent numerically insignificant terms involving the Wilson coefficients 
$c_3, \dots, c_6$ (see, e.g., \rf{fklms:exc}). 
The corresponding matrix element is 
\bea\label{mael:short}
\lefteqn{{\cal M}_{\sd}= \frac{G_F \a}{\sqrt{2} \p}V_{tb}^{}V_{tf}^{\ast}
\Bigg\{\bigg[(\ceff-c_{10}) \braket{P(k_1)P'(k_2)}{\bar{f}\g_{\m}P_L b}{\bar{B}(p)}}\nnu\\
&&\mbox{}-\frac{2\cseff}{q^2}\braket{P(k_1) P'(k_2)}{\bar{f}i \s_{\m\n}q^{\n}\left(m_b P_R+ m_f P_L\right) b}{\bar{B}(p)}\bigg]\bar{l}\g^{\m} P_L l + \bigg[c_{10}\to -c_{10}\bigg] \bar{l}\g^{\m} P_R l\Bigg\}.
\eea
Observe that the coefficient $\ceff$ contains both a weak phase (associated with the imaginary part of $\l_u$) and a strong phase 
[connected with the imaginary part of $g(m_i,s_l)$], thus opening the way to $\cp$-violating 
asymmetries between $B$ and $\bar{B}$ decays.

The hadronic part of the matrix element, describing the transition $\bar{B}\to PP'$, can be written in terms of $B\to V$ form factors ($V=K^*$ or $\r$)
\bea\label{formfactors:gmu}
\lefteqn{\braket{P(k_1)P'(k_2)}{\bar{f}\g_{\m}P_L b}{\bar{B}(p)}}\nnu \\ 
&&= -D_V(k^2)\Bigg\{ i\epsi_{\m\n\a\b}K^{\n}
p^{\a}k^{\b} g(q^2) +  \frac{1}{2}K_{\m} f(q^2) + k_{\m}\bigg[ (q\cdot W) a_+(q^2)-\frac{1}{2}\xi f(q^2)\bigg] 
+ \cdots \Bigg\},
\eea  
with the convention  $\epsi_{0123} = +1$,  
\bea\label{formfactors:smunu}
\lefteqn{\braket{P(k_1) P'(k_2)}{\bar{f}i \s_{\m\n}q^{\n}P_{R,L} b}{\bar{B}(p)}}\nnu\\
&&=D_V(k^2)\Bigg\{ -i\epsi_{\m\n\a\b}K^{\n}p^{\a}k^{\b}g_+(q^2)
\mp\frac{1}{2}K_{\m}\bigg[g_+(q^2) \D + q^2 g_-(q^2)\bigg]\nnu\\
&&\phantom{=\s\Bigg\{ }
\pm  k_{\m}\bigg\{(q\cdot W)\bigg[g_+(q^2) + \frac{1}{2}q^2 h(q^2)\bigg]
+ \frac{1}{2}\xi\bigg[g_+(q^2)\D + q^2 g_-(q^2)\bigg]\bigg\}+\cdots \Bigg\},
\eea
where the ellipses denote terms proportional to $q_{\m}$ which may be dropped in the case of massless leptons and
\be\label{def:delta}
\D = (M_B^2-M_V^2), \quad W^{\m} = K^{\m}-\xi k^{\m}, \quad
\xi = \frac{M_P^2-M_{P'}^2}{k^2}.
\ee
Throughout this paper,  we neglect the lepton mass, and assume the above form 
factors to be real, the absorptive parts due to real $c\bar{c}$ and $u\bar{u}$ 
states having been included in the functions $g(m_c,s_l)$ and $g(m_u, s_l)$ 
appearing in the short-distance coefficient $\ceff$ [\eq{wilsonc9}].
Furthermore,  we will limit ourselves to neutral $B$ mesons, i.e. 
$\bar{B}\equiv \bar{B}^0$.
  
The  function  $D_V(k^2)$ appearing in  Eqs. (\ref{formfactors:gmu}) and (\ref{formfactors:smunu}) 
is defined via
\be\label{def:sigma}
|D_V(k^2)|^2  = \frac{48 \p^2}{\b^3 M_V^2} \d(k^2-M_V^2),
\ee
where we have used a narrow-width approximation, and 
\be\label{def:beta}
\b = \frac{\l^{1/2}(k^2, M_P^2, M_{P'}^2)}{k^2},
\ee
with the triangle function 
\be\label{def:triangle}
\l (a,b,c) = a^2 + b^2 + c^2 - 2 (a b + b c + a c).
\ee
We adopt the $B\to K^*$ and $B\to \rho$ form factors $g$, $f$, $h$,  
$g_{\pm}$, $a_+$ given by Melikhov and Nikitin \cite{melikhov} 
(see Appendix \ref{formfactorsMel}). The matrix element (\ref{mael:short}) can then 
be written compactly as
\be\label{SD-matrix}
{\cal M}_{\sd}= \frac{G_F\a}{\sqrt{2}\p} V_{tb}^{}V_{tf}^{\ast}\bigg\{
( i \epsi_{\m\n\a\b}K^{\n} k^{\a} q^{\b} x_L+ K_{\m}y_L + k_{\m} z_L)\bar{l}\g^{\m}P_L l + (L\to R)\bigg\},
\ee
where
\be\label{shorthand:x}
x_{L, R}  = D_V(k^2)\bigg\{ (\ceff \mp c_{10})g(q^2) - \frac{2\cseff}{q^2} (m_b+ m_f)g_+(q^2)\bigg\}, 
\ee
\be\label{shorthand:y}
y_{L, R}  =  -\frac{1}{2}D_V(k^2) \bigg\{(\ceff \mp c_{10})f(q^2) 
- \frac{2\cseff }{q^2}(m_b-m_f) [g_+(q^2)\D + q^2g_-(q^2)]\bigg\}, 
\ee
\be\label{shorthand:z}
z_{L, R} = (q\cdot W) z'_{L, R} - \xi y_{L, R},
\ee
with
\be
z'_{L, R}  =  -D_V(k^2) \bigg\{(\ceff \mp c_{10})a_+(q^2) + \frac{2\cseff}{q^2}(m_b-m_f)\bigg[g_+(q^2) + \frac{1}{2}q^2 h(q^2)\bigg]\bigg\}.
\ee
The matrix element for the corresponding antiparticle channel can be 
obtained from (\ref{SD-matrix}) by means of CPT invariance. In fact, we have
\be\label{cpt:1}
\ol{\cal M}_{\sd} = 
{\cal M}_{\sd}(x_{L, R}\to -\bar{x}_{L, R}; y_{L, R}\to 
\bar{y}_{L, R}; z_{L, R}\to \bar{z}_{L, R})
\ee
for the conjugate processes $\bkc$ and $\bpc$, 
where $\bar{x},  \bar{y}, \bar{z}$ are related to $x, y, z$ by 
\be\label{cpt:2}
\phi_{\weak}\to -\phi_{\weak}, \quad \d\to \d.
\ee
Here  $\phi_{\weak}$ and $\d$ denote  the weak and strong phases respectively that appear in the matrix element ${\cal M}_{\sd}$. 

\section{Differential decay rate}
Squaring the matrix element (\ref{SD-matrix}) and summing over spins, 
we obtain for the decay $\bar{B}\to P P' l^-l^+$ \cite{nita:rahul}
\bea\label{mael:squared}
\lefteqn{\left|{\cal M}_{\sd}(\bar{B}\to P P' l^- l^+)\right|^2
=\frac{G_F^2\a^2 }{2\p^2}\,|V_{tb}^{}V_{tf}^{\ast}|^2}\nnu\\
&\times&\bigg\{2\Re (y_{L}^*z_{L}+y_{R}^*z_{R}) \Big[ (k\cdot q) (q\cdot K) -
     (k\cdot Q) (K\cdot Q) -s_l(k\cdot K)\Big]\nnu \\ [.7ex]
&+&2 \Re (x _{L}^* z_{L} - x _{R}^*z_{R})
  \Big[ (k\cdot q) (k\cdot Q) (q\cdot K)+ s_l s_P (K\cdot Q) -
     (k\cdot q)^2 (K\cdot Q)\nnu\\ 
&&\phantom{2 \Re (z_{L}x _{L}^* - z_{R}x _{R}^*)\Big[}
-s_l(k\cdot K)(k\cdot Q)\Big]\nnu \\[.7ex] 
 &+&2\Re (x _{R}y _{R}^* -x _{L}y _{L}^*)\Big[ s_l K^2 (k\cdot Q) - (k\cdot Q) (q\cdot K)^2 
+ (k\cdot q) (q\cdot K) (K\cdot Q)\nnu\\
&&\phantom{2\Re (x _{R}y _{R}^* -x _{L}y _{L}^*)\Big[}
-s_l(k\cdot K)(K\cdot Q) \Big]\nnu\\ [.7ex]
&+& ( |x_L|^2 + |x_R|^2) \Big[- 2 (k\cdot q) (k\cdot Q) (q\cdot K) (K\cdot Q) - s_l s_P (K\cdot Q)^2 +
     (k\cdot q)^2 (K\cdot Q)^2 \nnu \\  
&&\phantom{( x_L^2 + x_R^2) \Big[} - s_l K^2  (k\cdot Q)^2  + (k\cdot Q)^2 (q\cdot K)^2 
+2s_l (k\cdot K)(k\cdot Q)(K\cdot Q)\Big] \nnu\\ [.7ex]
&+& ( |y_L|^2 + |y_R|^2)\Big[ -s_l K^2 + (q\cdot K)^2 - (K\cdot Q)^2 \Big] \nnu\\[.7ex]
&+& (|z_L|^2 + |z_R|^2 ) \Big[ - s_l s_P  + (k\cdot q)^2 - (k\cdot Q)^2 \Big]\nnu\\[.7ex]
&+& 2\epsi_{\m\n\a\b}k^\m K^\n q^\a Q^\b
   \Big[ (K\cdot Q)  \Im (x _{L} y_{L}^* + x _{R} y _{R}^*)+
   \Im (y _{R}^* z_{R} -y _{L}^* z_{L})\nnu\\
&&\hspace{3cm}\mbox{} - (k\cdot Q) \Im (x_{L}^* z_{L}+x_{R}^* z_{R}) \Big]\bigg\},
\eea
with $s_P\equiv k^2$.
Introducing the shorthand notation
\be\label{def:X}
X = \Big[(k\cdot q)^2 - s_ls_P\Big]^{1/2} = \frac{1}{2}\l^{1/2}(M_B^2, s_l, s_P),
\ee
we find that ($m_l = 0$)
\begin{mathletters}
\be
k\cdot K = M_P^2-M_{P'}^2= \xi s_P,
\ee
\be
k\cdot q = \frac{1}{2}(M_B^2 - s_l - s_P),
\ee
\be
k\cdot Q = X \cos\theta_l,
\ee
\be
q\cdot K = \b X \cos\theta_P +  \xi (k\cdot q),
\ee
\be\label{product:qW}
q\cdot W = \b X \cos\theta_P, 
\ee
\be
K\cdot Q = \xi (k\cdot Q)  + \b\Big[ (k\cdot q)\cos\theta_l\cos\theta_P - (s_l s_P)^{1/2} \sin\theta_l\sin\theta_P\cos\phi\Big],
\ee
\be
\epsi_{\m\n\a\b}k^\m K^\n q^\a Q^\b = - (s_ls_P)^{1/2} \b X \sin\theta_l\sin\theta_P\sin\phi,
\ee
\be
K^2 = 2(M_P^2 +M_{P'}^2 )  - s_P= (\xi^2-\b^2)s_P, \quad Q^2 = - s_l.
\ee
\end{mathletters}%
To write the differential decay rate in compact form, we define the 
auxiliary functions
\be\label{funcs:f1}
F_{1L, R} = \b [X^2 z_{L, R}'  + (k\cdot q)y_{L, R}], 
\ee
\be\label{funcs:f2}
F_{2L, R} = \b(s_l s_P)^{1/2} y_{L, R},
\ee
\be\label{funcs:f3}
F_{3L, R} = \b X (s_ls_P)^{1/2} x_{L, R},
\ee
so that
\bea\label{diff:fivefold}
d^5\G= \frac{G_F^2 \a^2}{2^{16}\p^8M_B^3}\,|V_{tb}^{}V_{tf}^{\ast}|^2 \b X 
I(s_l, s_P, \theta_l, \theta_P, \phi)ds_l\,  ds_P\,  d\cos\theta_l\,  d\cos\theta_P\, d\phi,
\eea
with
\bea\label{funcs:i}
I &=& I_1 + I_2\cos 2\theta_l + I_3 \sin^2\theta_l\cos 2\phi + I_4 \sin 2\theta_l \cos\phi + I_5 \sin\theta_l\cos\phi 
+ I_6 \cos\theta _l \nnu\\
&&\mbox{} +  I_7 \sin\theta_l\sin\phi + I_8 \sin 2\theta_l \sin\phi + I_9 \sin^2\theta_l\sin 2\phi.
\eea
The functions $I_1, \dots, I_9$ have the following form (see 
Appendix \ref{funcslist:i} for details): 
\begin{mathletters}\label{Isubis}
\bea
I_1 &=& \left[ \frac{3}{2}(|F_{3L}|^2 + |F_{2L}|^2 )\sin^2\theta_P+ 
|F_{1L}|^2\cos^2\theta_P\right] + (L\to R),
\eea
\bea
I_2 &=& \left[ \frac{1}{2}( |F_{3L}|^2+ |F_{2L}|^2  )\sin^2\theta_P-|F_{1L}|^2\cos^2\theta_P \right] + (L\to R),
\eea
\bea
I_3 = (|F_{3L}|^2 - |F_{2L}|^2 )\sin^2\theta_P + (L\to R),
\eea
\bea
I_4 = \Re (F_{1L}^{}F^*_{2L}) \sin 2\theta_P + (L\to R),
\eea
\bea
I_5 = 2 \Re (F_{1L}^{}F^*_{3L}) \sin2\theta_P - (L\to R),
\eea
\bea
I_6 = 4\Re (F_{2L}^{}F^*_{3L}) \sin^2\theta_P - (L\to R),
\eea
\bea
I_7 =  2 \Im (F_{1L}^{}F^*_{2L}) \sin2\theta_P - (L\to R),
\eea
\bea
I_8 =  \Im (F_{1L}^{}F^*_{3L}) \sin2\theta_P + (L\to R),
\eea
\bea
I_9 = -2 \Im (F_{2L}^{}F^*_{3L}) \sin^2\theta_P + (L\to R).
\eea
\end{mathletters}%
The physical region of phase space is defined through
\begin{mathletters}\label{int:region}
\be
0\leqslant s_l\leqslant (M_B-\sqrt{s_P})^2, \quad (M_P + M_{P'})^2\leqslant s_P\leqslant M_B^2, 
\ee
\be
0\leqslant\phi\leqslant 2\p, \quad -1\leqslant\cos\theta_P\leqslant 1, \quad -1\leqslant\cos\theta_l\leqslant 1.
\ee 
\end{mathletters}%
Note that $\theta_l$ is the angle of the $l^-$ in the $l^-l^+$ c.m.~frame; $\theta_P$ is the angle of $\p^-$ (or $K^-$) in the $\p^-\p^+$ (or $K^-\p^+$) system; $\phi$ is the angle between 
$\Vec{p}_{\p^-}\times\Vec{p}_{\p^+}$ and $\Vec{p}_{l^-}\times\Vec{p}_{l^+}$ (in the case of the $\p^-\p^+$ final state) and between $\Vec{p}_{K^-}\times\Vec{p}_{\p^+}$ 
and  $\Vec{p}_{l^-}\times\Vec{p}_{l^+}$ (in the case of the $K^-\p^+$ final state), 
$\Vec{p}_i$ denoting the 3-momentum vectors of the corresponding particles in the $\bar{B}$ rest frame. The $z$-axis is chosen along the total momentum vector of the $PP'$ system in the $\bar{B}$ rest frame.
This definition will also be retained in the case of the antiparticle reactions $\bpc$ and $\bkc$ (with $\Vec{p}_{K^+}$ replacing $\Vec{p}_{K^-}$ in the latter case).    

The amplitudes $F_{iL, R}$ ($i=1,2,3$) defined in \eqs{funcs:f1}{funcs:f3} are closely related to the transversity amplitudes 
$A_0, A_{\|}, A_{\bot}$ sometimes used in connection with the angular 
distribution of the four-body final state arising from decays of the form 
$B\to V_1V_2$ \cite{btovecvec}. The differential decay rate in this alternative notation is again given by \eqs{diff:fivefold}{Isubis}, with the functions 
$F_{iL, R}$ written in terms of $A_0, A_{\|}, A_{\bot}$ as follows:
\begin{mathletters}
\be
F_{1L, R}=\frac{A_{0L,R}}{N},
\ee
\be
F_{2L, R}=\frac{A_{\|L,R}}{N \sqrt{2}},
\ee
\be
F_{3L, R}=\frac{A_{\bot L,R}}{N \sqrt{2}},
\ee
\end{mathletters}
where the normalization factor
\be
N=\frac{1}{3}\Bigg[\frac{G_F^2 \a^2}{2^{11}\p^7M_B^3}\,|V_{tb}^{}V_{tf}^{\ast}|^2 \b X\Bigg]^{1/2}, 
\ee
has been chosen in such a way that $\G=|A_0|^2 + |A_{\|}|^2+
|A_{\bot}|^2$.

\section{Angular distributions}
We derive from the differential decay rate, \eq{diff:fivefold}, 
one-dimensional angular distributions of interest, namely $d\G/d\cos\theta_P$, $d\G/d\cos\theta_l$, and $d\G/d\phi$.
These distributions, as well as the observables calculated in \sec{cp},
depend on different combinations of the short-distance coefficients $\cseff$, 
$\ceff$, $c_{10}$ and the form factors $g$, $f$, $h$, $g_{\pm}$, $a_+$.

\subsection{Decay rate as a function of \bm$\cos\theta_P$}  
Integrating $d^5\G$ over the variables $s_l$, $s_P$, $\cos\theta_l$, 
and $\phi$, we obtain
\be\label{distr:thetaP}
\frac{d\G}{d\cos\theta_P}= \frac{G_F^2M_B^5\a^2}{2^{14}\p^7}\,
|V_{tb}^{}V_{tf}^{*}|^2\left(J_1 - \frac{1}{3}J_2\right),
\ee
with
\be\label{def:J}
J_i\equiv J_i^s\sin^2\theta_P + J_i^c\cos^2\theta_P, 
\quad J_i^{s,c}= \frac{1}{M_B^8}\int I_i^{s,c}\b X ds_l\, ds_P,\quad i=1,2,
\ee
where $I_i^{s,c}$ stand for the coefficients of $\sin^2\theta_P$ and 
$\cos^2\theta_P$ in the expressions for $I_{1,2}$ given in 
Eqs.~(\ref{def:I1}) and (\ref{def:I2}) of the Appendix.
The absence of a term odd in $\cos\theta_P$ is connected with the fact that the $PP'$ system is in a pure $L=1$ state.
As a consequence, the forward-backward (FB) asymmetry in the $PP'$ system, 
defined as 
\be\label{FB}  
A_{\mathrm{FB}}^P=\frac{\dis \int_0^1 \frac{d\G}{d\cos\theta_P}d\cos\theta_P-
\int_{-1}^0 \frac{d\G}{d\cos\theta_P}d\cos\theta_P}{\dis \int_0^1 
\frac{d\G}{d\cos\theta_P}d\cos\theta_P+\int_{-1}^0\frac{d\G}{d\cos\theta_P}
d\cos\theta_P}, 
\ee
vanishes. The predictions for $J_i^{s,c}$ are contained in Table \ref{tableJ}.
%
%
\begin{table}
\caption{Values of the functions $J_i^{s,c}$ [\eq{def:J}],  
using $\rho=0.19$ and $\eta=0.35$ \protect\cite{ckm:analysis}.\label{tableJ}}
\begin{tabular}{ccccc}
$ $& $J_1^s$& $J_1^c$&$J_2^s$& $J_2^c$ \\
\hline
$\bp$ &$76.2$& $76.8$& $25.4$ &  $-76.8$\\
\hline
$\bk$ &$261.4$& $278.8$& $87.2$ & $-278.8$
\end{tabular}
\end{table}
%
%
\subsection{Decay rate as a function of \bm$\cos\theta_l$} 
Integration of $d^5\G$ over $s_l$, $s_P$, $\cos\theta_P$, and $\phi$ yields
\be\label{distr:thetal}
\frac{d\G}{d\cos\theta_l}= \frac{G_F^2M_B^5\a^2}{2^{15}\p^7}\,
|V_{tb}^{}V_{tf}^{*}|^2
(K_1 + K_2\cos2\theta_l + K_6 \cos\theta_l),
\ee
where
\be\label{Kis}
K_i=\frac{1}{M_B^8}\int I_i \b X ds_l\,ds_Pd\cos\theta_P.
\ee
Our results for the parameters $K_i$ are tabulated in Table \ref{tableK}.
Observe that $K_4$, $K_5$, $K_7$, and $K_8$ vanish in our model.
%
%
\begin{table}
\caption{Estimate of the functions $K_i$ according to \eq{Kis}.}\label{tableK}
\begin{tabular}{cccccccccc}
$ $& $K_1$& $K_2$&$K_3$& $K_4$& $K_5$ & $K_6$ & $K_7$ & $K_8$& $K_9$\\
\hline
$\bp$ &$152.8$& $-17.3$& $-14.1$ &  $0$& $0$ & $-61.8$& $0$& 
$0$ & $-0.05$\\
\hline
$\bk$ &$534.4$& $-69.6$& $-64.6$ & $0$& $0$& $-179.0$ &$0$& 
$0$ & $-0.4$
\end{tabular}
\end{table}
%
%
The FB asymmetry of $l^-$ in the $l^-l^+$ c.m. system is
\be\label{asym:FB}
A_{\mathrm{FB}} = \frac{K_6/2}{K_1-K_2/3}= \left\{\begin{array}{l}- 0.19\quad \text{for}\quad \p^-\p^+,\\
-0.16 \quad \text{for}\quad K^-\p^+.
\end{array}\right. 
\ee
This result reproduces the FB asymmetry of the lepton calculated in previous analyses of the exclusive channels $\bar{B}\to K^* (\r) e^-e^+$ 
\cite{lit:btokstar,fklms:exc}.

\subsection{Decay rate as a function of \bm$\phi$}
Finally, the distribution in the angle $\phi$ between the lepton and meson planes, after integration over other variables, takes the simple form
\be
\frac{d\G}{d\phi}= \frac{G_F^2M_B^5\a^2}{2^{15}\p^8}\,
|V_{tb}^{}V_{tf}^{*}|^2\Bigg[(K_1 -\frac{1}{3}K_2) + \frac{2}{3}(K_3 \cos 2\phi + K_9 \sin2 \phi)\Bigg],
\ee
with the $K_i$'s shown in Table \ref{tableK}. 

\section{$\bbox{\cp}$-violating observables}\label{cp}
We now focus on $\cp$-violating observables that can be constructed by 
combining information on $\bar{B}$ and $B$ decays, namely
\begin{mathletters}
\be
\bk\quad \text{and}\quad \bkc,
\ee
or
\be
\bp\quad \text{and}\quad \bpc.
\ee
\end{mathletters}
With the definition of $\cos\theta_P$, $\cos\theta_l$, and $\phi$ given after \eq{int:region},
the differential decay rate for $\bar{B}$ and $B$ decays is
\begin{mathletters}\label{dgamma}
\be
d^5\G|_{\bar{B}}= {\mathcal{N}} I(s_l, s_P, \theta_l, \theta_P, \phi)\b X {d\mathrm{PS}},
\ee
\be
d^5\G|_B= {\mathcal{N}} \bar{I}(s_l, s_P, \theta_l, \theta_P, \phi)\b X {d\mathrm{PS}},
\ee
\end{mathletters}%
where $\mathcal{N}$ is a normalization factor
\be
{\mathcal{N}}=\frac{G_F^2\a^2}{2^{16}\p^8M_B^3}\,|V_{tb}^{}V_{tf}^{*}|^2,
\ee
and $d\mathrm{PS}$ represents the phase-space element
\be
{d\mathrm{PS}}= ds_l\, ds_P\,d\cos\theta_l\, d\cos\theta_P\, d\phi.
\ee
The function $I$ is given in \eq{funcs:i}, whereas the function $\bar{I}$ is obtained from $I$ by the substitution
\begin{mathletters}
\be
I_{1,2,3,4,7}(x_{L,R};y_{L,R};z_{L,R})\longrightarrow
I_{1,2,3,4,7}(\bar{x}_{L,R};\bar{y}_{L,R};\bar{z}_{L,R})\equiv 
\bar{I}_{1,2,3,4,7},
\ee
\be
I_{5,6,8,9}(x_{L,R};y_{L,R};z_{L,R})\longrightarrow
- I_{5,6,8,9}(\bar{x}_{L,R};\bar{y}_{L,R};\bar{z}_{L,R})\equiv 
- \bar{I}_{5,6,8,9},
\ee
\end{mathletters}%
where $\bar{x}$, $\bar{y}$, $\bar{z}$ are defined via \eq{cpt:2}.

We now define the sum and the difference of the differential spectra 
$d^5\G|_{\bar{B}}$ and $d^5\G|_B$ as follows:
\bea
d\G_{\text{diff}}&=& {\mathcal{N}} \bigg[(I_1-\bar{I}_1)+(I_2-\bar{I}_2)\cos 2\theta_l
+ (I_3-\bar{I}_3)\sin^2\theta_l\cos 2\phi + (I_4-\bar{I}_4)\sin2\theta_l\cos\phi
\nnu\\
&+& (I_5+\bar{I}_5)\sin\theta_l\cos\phi+ (I_6+\bar{I}_6)\cos\theta_l+ (I_7-\bar{I}_7)
\sin\theta_l\sin\phi +(I_8+\bar{I}_8)\sin2\theta_l\sin\phi\nnu\\
&+& (I_9+\bar{I}_9)\sin^2\theta_l\sin 2\phi\bigg]
\b X d{\mathrm{PS}},
\eea  
\bea
d\G_{\text{sum}}&=& {\mathcal{N}} \bigg[(I_1+\bar{I}_1)+(I_2+\bar{I}_2)\cos 2\theta_l
+ (I_3+\bar{I}_3)\sin^2\theta_l\cos 2\phi + (I_4+\bar{I}_4)\sin2\theta_l\cos\phi
\nnu\\
&+& (I_5-\bar{I}_5)\sin\theta_l\cos\phi+ (I_6-\bar{I}_6)\cos\theta_l+ (I_7+\bar{I}_7)
\sin\theta_l\sin\phi +(I_8-\bar{I}_8)\sin2\theta_l\sin\phi\nnu\\
&+& (I_9-\bar{I}_9)\sin^2\theta_l\sin 2\phi\bigg]
\b X d{\mathrm{PS}}.
\eea

Recalling \eq{Kis} and defining
\be
K_i^D=\frac{1}{M_B^8}\int \b X ds_l\,ds_P \Bigg[\int_{-1}^0-\int_0^1\Bigg] 
I_i d\cos\theta_P , 
\ee
we can then construct the following $\cp$-violating observables:
\begin{mathletters}\label{def:cpasym}
\be
\av{A_{\mathrm{CP}}}= \frac{1}{I_0}\bigg[(K_1-\bar{K}_1) - 
\frac{1}{3}(K_2-\bar{K}_2)\bigg], 
\ee
\be
\av{A_3}= \frac{1}{I_0}(K_3-\bar{K}_3),
\ee
\be
\av{A_4}= \frac{1}{I_0}(K_4^D-\bar{K}_4^D),
\ee
\be
\av{A_5}= \frac{1}{I_0}(K_5^D-\bar{K}_5^D),
\ee
\be
\av{A_6}= \frac{1}{I_0}(K_6-\bar{K}_6),
\ee
\be
\av{A_7}= \frac{1}{I_0}(K_7^D-\bar{K}_7^D),
\ee
\be
\av{A_8}= \frac{1}{I_0}(K_8^D-\bar{K}_8^D),
\ee
\be
\av{A_9}= \frac{1}{I_0}(K_9-\bar{K}_9),
\ee
\end{mathletters}%
in which the quantity $I_0$ is defined as
\be
I_0=(K_1+\bar{K}_1) - \frac{1}{3}(K_2+\bar{K}_2).
\ee 
Note that the asymmetries $\av{A_{\mathrm{CP}}}$, $\av{A_{3,4,7}}$ 
involving the differences $(I_i-\bar{I}_i)$, $i=1,2,3,4,7$, 
can be obtained from a measurement of the difference $d\G_{\mathrm{diff}}$. 
On the other hand, the asymmetries $\av{A_{5,6,8,9}}$ require only a 
measurement of $d\G_{\mathrm{sum}}$ and, in principle, can be
determined even for an untagged equal mixture of $B$ and $\bar{B}$.

The asymmetry $\av{A_{\mathrm{CP}}}$ is simply the asymmetry in the partial decay rates of 
$\bk$ and $\bkc$ (or $\bp$ and $\bpc$). The other asymmetries represent $\cp$-violating effects in the angular distribution of these processes. All of these effects have their origin in the $\cp$-violating imaginary part of the coefficient
\be
\l_u \equiv \frac{V_{ub}^{}V_{uf}^*}{V_{tb}^{}V_{tf}^*}\approx 
\left\{\begin{array}{l}-\l^2(\r-i\eta) \quad \text{for}\quad f=s ,\\
\dis\frac{\rho(1-\rho)-\eta^2}{(1-\rho)^2 + \eta^2} 
- \frac{i \eta}{(1-\rho)^2 + \eta^2} \quad \text{for}\quad f=d .
\end{array}\right.
\ee
In the case $f=d$, this imaginary part is of order $\eta$, but in the case
$f=s$  it is reduced by an extra factor $\l^2$. In both cases, the weak phase 
present in the coefficient $\ceff$ [\eq{wilsonc9}] is further suppressed by a factor of order $(3c_1 + c_2)/c_9\simeq 0.085$. All of these factors explain the very small magnitude of the asymmetries listed in Table \ref{tableAsym}. 
%
%
\begin{table}
\caption{Estimates of the average $\cp$-violating asymmetries 
$\av{A_k}$ in units of $10^{-4}$ ($10^{-2}$) for the $B\to K^*$ ($B\to \r$) transition.}\label{tableAsym}
\begin{tabular}{ccccccccc}
& $\av{A_{\mathrm{CP}}}$& $\av{A_3}$& $\av{A_4}$&$\av{A_5}$& $\av{A_6}$& $\av{A_7}$ & $\av{A_8}$& $\av{A_9}$\\
\hline
$\begin{array}{c}\bk\\[-2ex]\mathrm{vs}\\[-2ex]\bkc\end{array}$ &
$2.7$ & $-0.6$ & $-2.0$ & $5.2$ &  $-4.6$ & $0$ & $0.6$ & $-0.04$\\
\hline
$\begin{array}{c}\bp\\[-2ex] \mathrm{vs}\\[-2ex] \bpc\end{array}$ &
$-1.7$ & $0.1$ & $0.4$ & $-1.4$ & $1.2$  & $0$ & $-0.1$ & $0.006$
\end{tabular}
\end{table}
%
%
The result for the partial width asymmetry $\av{A_{\mathrm{CP}}}$ reproduces 
that obtained  in previous literature \cite{fklms:exc}.

The small value of $\av{A_6}$ implies that the FB asymmetry of the 
electron  is opposite in sign for $B$ and $\bar{B}$ decay
[\eq{asym:FB}]. This result agrees with the statement in \rf{talk:lms}.

\section{Conclusions}
Our results for $d\G/d \cos\theta_P$ and $d\G/d\phi$ are new consequences of the short-distance Hamiltonian for $b\to s(d) e^-e^+$, which test the polarization state of the vector meson in $\bar{B}\to K^* (\r) e^-e^+$. The asymmetries $\av{A_k}$ are $\cp$-violating observables that can be obtained from a comparison of the angular distribution in $B$ and $\bar{B}$ decays in the conjugate channels. Numerical estimates have been obtained within the framework of the 
standard model, but the formalism 
presented can be applied to other Hamiltonians. 
As seen from Table \ref{tableAsym}, the asymmetries predicted for 
$B\to K^*$ are 
exceedingly small, and any significant effect would signal a non-standard 
source of $\cp$ violation. In the case of $B\to \rho$, asymmetries of $1$--$2\%$ are predicted for $\av{A_{\mathrm{CP}}}$, $\av{A_5}$, and $\av{A_6}$.  
It should be noted that our predictions apply to $K^*$ ($\r$) production on the mass shell. 
This implies that the $PP'$ system is in a pure $p$-wave state, and explains 
why the distribution in $\cos\theta_P$ given in \eq{distr:thetaP} 
is forward-backward sym\-me\-tric.
In the continuum region of $K\p$ or $\p\p$ masses, there will be  additional partial waves, as well as long-range effects associated with bremsstrahlung from $B\to K\p$ and $B\to \p\p$, followed by internal conversion of the photon. While such effects are difficult to calculate, it is conceivable that they lead to larger $\cp$-violating asymmetries in the continuum region of $B\to K\p e\bar{e}$ and $B\to \p\p e\bar{e}$.

\acknowledgments
The work of F.\,K. has been supported by the TMR Network of the EC under 
contract ERBFMRX-CT96-0090.

\appendix
\section{Form factors}\label{formfactorsMel}
In this Appendix we list the $B\to K^*$ and $B\to \rho$ form factors of 
Melikhov and Nikitin \cite{melikhov} using their ``Set 2'' of the 
Isgur-Scora-Grinstein-Wise  parameters  \cite{isguretal}. 
(See Table \ref{tableform}.)
%
%
\begin{table}[h]
\caption{The $B\to K^*$ and $B\to \rho$ form factors of Melikhov and Nikitin 
\protect\cite{melikhov}.}\label{tableform}
\begin{tabular}{lcc}
Form factors & $B\to K^*$ & $B^-\to \rho^-$  \\ \hline
\\[-4.5ex] 
$g(q^2)$ & $0.048\ \GeV^{-1}\left(1-\dis\frac{q^2}{6.67^2}\right)^{-2.61}$
& $0.036\ \GeV^{-1}\left(1-\dis\frac{q^2}{6.55^2}\right)^{-2.75}$ \\
\\[-4.5ex] 
$f(q^2)$ & $1.61\ \GeV\left(1-\dis\frac{q^2}{5.86^2}+\dis\frac{q^4}{7.66^4}\right)^{-1}$ 
& $1.10\ \GeV\left(1-\dis\frac{q^2}{5.59^2}+\dis\frac{q^4}{7.10^4}\right)^{-1}$ \\
\\[-4.5ex] 
$a_+(q^2)$ & $-0.036\ \GeV^{-1}\left(1-\dis\frac{q^2}{7.33^2}\right)^{-2.85}$ 
& $-0.026\ \GeV^{-1}\left(1-\dis\frac{q^2}{7.29^2}\right)^{-3.04}$ \\
\\[-4.5ex] 
$h(q^2)$ & $0.0037\ \GeV^{-2}\left(1-\dis\frac{q^2}{6.57^2}\right)^{-3.28}$
& $0.003\ \GeV^{-2}\left(1-\dis\frac{q^2}{6.43^2}\right)^{-3.42}$ \\
\\[-4.5ex] 
$g_+(q^2)$ & $-0.28\left(1-\dis\frac{q^2}{6.67^2}\right)^{-2.62}$
& $-0.20\left(1-\dis\frac{q^2}{6.57^2}\right)^{-2.76}$ \\
\\[-4.5ex] 
$g_-(q^2)$ & $0.24\left(1-\dis\frac{q^2}{6.59^2}\right)^{-2.58}$
& $0.18\left(1-\dis\frac{q^2}{6.50^2}\right)^{-2.73}$
\end{tabular}
\end{table}
%
%

\section{The functions \bm$I_1, \dots, I_9$}\label{funcslist:i}
The functions $I_1, \dots, I_9$ appearing in the expression for $I(s_l, s_P, \theta_l, \theta_P, \phi)$, \eq{funcs:i}, are given by
\bea\label{def:I1}
I_1&=&\b^2\Bigg\{\frac{3}{2}s_ls_P\bigg[X^2 (|x_L|^2 + |x_R|^2 )+ 
(|y_L|^2 + |y_R|^2 )\bigg]\sin^2\theta_P + \bigg[(k \cdot q)^2(|y_L|^2 + |y_R|^2)\nnu\\
&&\mbox{}+ X^4(|z_L'|^2 +|z_R'|^2) + 2 X^2 (k\cdot q)\Re(y_L^*z_L' +y_R^*z_R')\bigg]\cos^2\theta_P\Bigg\},  
\eea
\bea\label{def:I2}
I_2&=&\b^2\Bigg\{\frac{1}{2}s_ls_P\bigg[X^2 (|x_L|^2 + |x_R|^2 )+ 
(|y_L|^2 + |y_R|^2 )\bigg]\sin^2\theta_P - \bigg[(k \cdot q)^2(|y_L|^2 + |y_R|^2)\nnu\\
&&\mbox{}+ X^4(|z_L'|^2 +|z_R'|^2) + 2 X^2 (k\cdot q)\Re(y_L^*z_L' +y_R^*z_R')\bigg]\cos^2\theta_P\Bigg\},  
\eea
\be
I_3 = \b^2 s_ls_P\bigg\{X^2 (|x_L|^2 + |x_R|^2 ) - (|y_L|^2 + |y_R|^2 )\bigg\}\sin^2\theta_P,
\ee 
\be
I_4= \b^2 (s_ls_P)^{1/2}\bigg\{X^2 \Re ( y_L^*z_L'+ y_R^*z_R')
+ (k\cdot q)(| y_L|^2 +  |y_R|^2)\bigg\}\sin 2\theta_P,
\ee
\be
I_5= 2\b^2 X (s_ls_P)^{1/2}\bigg\{X^2 \Re ( x_L^*z_L'- x_R^*z_R')
+ (k\cdot q) \Re ( x_L^* y_L - x_R^* y_R)\bigg\}\sin 2\theta_P,
\ee
\be
I_6 = 4\b^2 X s_ls_P \Re ( x_L^* y_L - x_R^* y_R)\sin^2\theta_P,
\ee
\be
I_7 = 2\b^2 X^2 (s_ls_P)^{1/2}\Im(y_L^* z_L' - y_R^*z_R')\sin 2\theta_P,
\ee
\be
I_8= \b^2 X(s_ls_P)^{1/2}\bigg\{ X^2\Im(x_L^*z_L' + x_R^*z_R')
+ (k\cdot q) \Im(x_L^* y_L + x_R^* y_R)\bigg\}\sin 2\theta_P,
\ee
\be\label{def:I9}
I_9 = -2 \b^2 X s_l s_P \Im (x_L^* y_L + x_R^* y_R)\sin^2\theta_P,
\ee
where $k\cdot q = (M_B^2 - s_l- s_P)/2$. 
%
%

%
\end{document}